\documentclass[twocolumn]{autart}

\usepackage{graphicx}
\usepackage{epstopdf}
\usepackage{amsfonts,latexsym,amsmath,amssymb}
\usepackage[mathscr]{eucal}
\usepackage{graphicx,color}

\newtheorem{theorem}{Theorem}

\newtheorem{proposition}{Proposition}
\newtheorem{corollary}{Corollary}
\newtheorem{definition}{Definition}

\newcommand{\sbm}[1]{\left[\begin{smallmatrix} #1
	\end{smallmatrix}\right]}

\newcommand{\bluff}{{\hbox{\raise 15pt \hbox{\hskip 0.5pt}}}}

\def\rline{{\mathbb R}}    

\newcommand{\rfb}[1]{\mbox{\rm
		(\ref{#1})}\ifx\undefined\stillediting\else:\fbox{$#1$}\fi}

\def\startmodif{\color{black}}
\def\stopmodif{\color{black}}

\begin{document}
	
	\begin{frontmatter}
		
		\title{Robustness Analysis of Systems' Safety through a New Notion of Input-to-State Safety   \thanksref{footnoteinfo}}
		
		\thanks[footnoteinfo]{The work of M.Z. Romdlony is supported by the Indonesian DIKTI scholarship program  and the work of B. Jayawardhana is supported by the EU Interreg IV SmartBot project and by the SNN project on the Region of Smart Factories. This paper was not presented at any IFAC meeting. Corresponding author M.~Z.~Romdlony. Tel. +31-503637156. }
		
		\author[1,2]{Muhammad Zakiyullah Romdlony} 
		\author[1]{Bayu Jayawardhana} 
		
		\address[1]{Engineering and Technology Institute Groningen, Faculty of Mathematics and Natural Science, University of Groningen, The Netherlands (e-mail: \{m.z.romdlony, b.jayawardhana\}@rug.nl)}
		\address[2]{School of Electrical Engineering, Telkom University, Indonesia (e-mail: zakiyullah@telkomuniversity.ac.id)}


		\begin{abstract}
        In this paper, we propose a new robustness notion that is applicable for certifying systems' safety with respect to external disturbance signals. The proposed input-to-state safety (ISSf) notion allows us to certify systems' safety in the presence of the disturbances which is analogous to the notion of input-to-state stability (ISS) for analyzing systems' stability. 
		\end{abstract}
		
	\end{frontmatter}
	
	\section{Introduction}
	
	With the advent of complex cyber-physical systems (CPS) and industrial internet-of-thing, the safety of integrated cyber-physical systems has become an important design feature that must be incorporated in all software levels \cite{Banerjee2011}.  
	In particular, this feature must also be present in the low-level control systems where both aspects of safety and stability are integrated in the control design for safety-critical systems, such as, biomedical devices, smart infrastructure systems, and smart energy systems.     
	
	For the past few years, a number of control design methods have been proposed in literature on the design of feedback controller that can guarantee both the safety and stability, simultaneously. To name a few, we refer interested readers to 
	\cite{Ames2014}, \cite{Xu2015}, \cite{Romdlony2015} and \cite{Romdlony2016a}. In \cite{Ames2014} and \cite{Xu2015}, the authors proposed an optimization problem, in the form of a quadratic programming, where both control Lyapunov and control Barrier inequalities are formulated in the constraints. The proposed method generalizes the well-known pointwise min-norm control method for designing a control law using control Lyapunov functions via an optimization problem \cite{Primbs1999}. It has been successfully implemented in the cruise control of autonomous vehicle as reported in \cite{Mehra2015}. Another direct approach is pursued in \cite{Romdlony2014a,Romdlony2016a} which is based on the direct merging of control Lyapunov function and control Barrier function. The merging process results in a control Lyapunov-Barrier function which can be used to stabilize the system with guaranteed safety by using Sontag's universal control law.            
		
	Despite the appealing idea in the aforementioned works for guaranteeing stability and safety, it remains unclear on how to analyze the robustness of the closed-loop system in the presence of external (disturbance) input signals.  
	There are many tools available for analyzing the robustness of systems' stability, including, $H_\infty$ and $L_2$-stability theories \cite{Schaft2000,Jayawardhana2009}, absolute stability theory \cite{Jayawardhana2011}, input-to-state stability (ISS) theory \cite{Sontag1996} and many others. However, analogous tools for systems' safety are still lacking which makes it difficult to carry out robustness analysis to  the aforementioned results that deal with the problem of stabilization with guaranteed safety.  
	
	The seminal work in \cite{Sontag1989,Sontag1996} on the characterization of input-to-state stability has been one of the most important tools in the stability analysis of nonlinear systems. It has allowed us to study stability of interconnected systems, to quantify systems' robustness with respect to external disturbances and to provide means for constructing a robustly stabilizing control law. The use of ISS Lyapunov function is crucial in all of these applications. In the following decade, the concept of ISS has been used and/or generalized in various directions with a commonality on the robustness analysis of systems' stability. However, safety and constraint aspects have not been considered in this framework. By considering the complement of the set of unsafe state, one might consider to apply recent generalization of ISS to the stability of invariant sets as in \cite{Angeli2015}. But it may not give us an insightful detail on the influence of external disturbance signals to the state of safety of the system. In this case, the resulting ISS inequality will only provide us information on the effect of external input to the systems' trajectory with respect to the complement set of unsafe state, but not on how far it is from being unsafe.

	In this paper, we propose a new notion of input-to-state safety which is an adaptation of ISS inequality to the systems' safety case. In particular, instead of the usual ISS inequality where the state trajectory $x(t)$ of the system can be bounded from above by a term that depends on initial condition and decays to zero and another term that depends on the $L^\infty$-norm of the external input signal $u(t)$, we look at the following inequality \startmodif
	\begin{equation}\label{ISSafety_ineq}
		\sigma(|x(t)|_{\mathcal D}) \geq \min\{\mu\left(|x(0)|_{\mathcal D},t\right),\delta\} - \phi\left(\|u(t)\|\right)	
	\end{equation}\stopmodif
	where $\mathcal D$ is the set of unsafe state, $|x|_{\mathcal D}$ denotes the distance of $x$ to $\mathcal D$, the function $\sigma$ is strictly increasing function,  $\mu$ is strictly increasing function in both arguments, $\delta>0$ and $\phi$ as the gain function that is dependent on input $u$, akin to the ISS case. As will be discussed later in Section 3, the inequality \rfb{ISSafety_ineq} will be called input-to-state safety (ISSf) inequality. 
	
	Roughly speaking, this inequality can be interpreted as follows. When there is no external input signal $u$, then the state trajectory will never get closer to $\mathcal D$. On the other hand, if there is an external input signal then it may jeopardize the systems' safety when the input signal $u$ is taken sufficiently large.
    
    The above interpretation serves very well with what we can expect in real systems where external disturbance input can potentially bring the system into the unsafe state. Xu {\it etal.} in \cite{Xu2015} has presented also a preliminary study on the robustness aspect for systems' safety where they provide an indirect relationship between the external input norm to the admissible initial conditions such that the system remains safe. This relationship is also captured in \rfb{ISSafety_ineq} where if the bound on the input signal is known then the inequality \rfb{ISSafety_ineq} will make sense only if the initial conditions are bounded away from $\mathcal D$ by a constant that depends on the input norm. 
		
	Complementary to the work of Xu {\it etal.} in \cite{Xu2015}, we adapt the ISS framework a'la Sontag to the systems' safety case through the use of ISSf barrier function which implies \rfb{ISSafety_ineq}. Preliminary work on this concept has been presented in \cite{Romdlony2016b} which is restricted to the case of exponential input-to-state safety. In this paper, we extend it to general nonlinear case, as well as to the analysis of feedback interconnection.

	This paper is organized as follows. In Section 2, we briefly recall the notion of stabilization with guaranteed safety, of ISS and of barrier certificate. 
	In Section 3, we introduce formally the notion of input-to-state safety and its characterization using ISSf barrier function.
   \startmodif In Section 4, we provide a numerical example of the aforementioned results for a simple mobile robot navigation system.  \stopmodif
	
	\section{Preliminaries}
	\textbf{Notation}. Throughout this paper, we consider an affine non-linear system described by
	\begin{equation}
		\label{NL}\dot{x} = f(x) + g(x)u, \qquad x(0)=x_{0},
	\end{equation}
	where $x(t) \in\mathbb{R}^{n}$ denotes a state vector,  $u(t)\in\mathcal{U}\subseteq\mathbb{R}^{m}$ denotes an (external) input or disturbance to the system. The functions $f(x)$ and $g(x)$ are
	$\mathcal{C}^1$ where the space $\mathcal C^1(\rline^l,\rline^m)$ consists of all continuously differentiable functions $F:\rline^l\to\rline^m$. \startmodif Without loss of generality and for simplicity of presentation, we will assume throughout that the solution to \rfb{NL} is complete (i.e., it exists for all $t\geq 0$) for any bounded signal $u$. This assumption holds when the system has the input-to-state stability property which we will recall shortly. \stopmodif 
	
	For a given signal $x:\rline_+\to\rline^n$, its $L^p$ norm is given by $\|x\|_{L^p} :=(\int^{\infty}_{0}\|x(t)\|^p dt)^{1/p}$ for $p=[1,\infty)$ and its $L^\infty$ norm is defined by $\|x\|_{L^\infty} := {\rm (ess) \ sup}_{t}(\|x(t)\|)$. For a given bounded set $\mathcal{M}\subset\mathcal{X}\subset\mathbb{R}^n$, we define the distance of a point $\xi\in\mathbb{R}^n$ with respect to $\mathcal{M}$ by $|\xi|_\mathcal{M}:=\min_{a\in\mathcal{M}}\|\xi-a\|$ where $\| \cdot \|$ is a metric norm. We define an open ball centered at a point $a \in \mathbb{R}^n$ with radius $r>0$ by $\mathbb{B}_r(a):=\{\xi\in \mathbb{R}^n|\|\xi-a\|<r\}$ and its closure is denoted by $\overline{\mathbb{B}}_r(a)$.
	
	We define the class of continuous strictly increasing functions $\alpha:\rline_+\to\rline_+$ by $\mathcal P$ and denote by $\mathcal{K}$ all functions $\alpha\in \mathcal P$ which satisfy 
	$\alpha(0)=0$. Moreover, $\mathcal{K}_\infty$ denotes all functions $\alpha\in\mathcal{K}$ which satisfy  $\alpha(r)\rightarrow\infty$ as $r\rightarrow\infty$.\startmodif By $\mathcal{KL}$ we denote all functions $\beta:\mathbb{R}_+\times\mathbb{R}_+\rightarrow\mathbb{R}_+$ such that $\beta(\cdot,t)\in\mathcal{K}$ for a fixed $t\geq 0$ and and $\beta(s,\cdot)$ is 
decreasing and converging to zero for a fixed $s\geq 0$.\stopmodif Correspondingly, we also denote by $\mathcal{KK}$ all functions $\mu:\mathbb{R}_+\times\mathbb{R}_+\rightarrow\mathbb{R}_+$ such that $f(0,0)=0$ and $f(s,t)$ is srictly increasing in both arguments.
	
	Let $\mathcal{X}_{0}\subset\mathbb{R}^{n}$ be the set of initial conditions and let an open and bounded set $\mathcal{D}\subset\mathbb{R}^{n}$ be the set of unsafe states, where we assume that $\mathcal{D}\cap\mathcal{X}_{0}=\emptyset$. 
    For a given set $\mathcal{D}\subset\mathbb{R}^{n}$, we denote the boundary of $\mathcal{D}$ by $\partial\mathcal{D}$ and the closure of $\mathcal{D}$ by $\overline{\mathcal{D}}$.
	
	Following safety definition in \cite{Romdlony2016a}, the (autonomous) system \rfb{NL} with $u=0$ is called \emph{safe} if for all $x_{0}\in\mathcal{X}_{0}$ and for all $t\in\overline{\mathbb{R}}_{+}$,  $x(t)\notin\overline{\mathcal{D}}$. Additionally, \rfb{NL} with $u=0$ is called (asymptotically) stable with guaranteed safety if it is both (asymptotically) stable and safe.  
	Based on these notions, the problem of stabilization with guaranteed safety has been investigated in \cite{Romdlony2016a} where the control problem is to 
	design a feedback law $u=k(x)$ such that the closed loop system is safe and asymptotically stable, i.e. for all $x_0\in\mathcal{X}_0$, we have that $x(t)\notin \mathcal{D}$ for all $t$ and $\displaystyle{\lim_{t\rightarrow \infty}\|x(t)\|=0}$. Moreover, when $\mathcal{X}_{0}=\mathbb{R}^{n}\setminus\mathcal{D}$ the problem is called {\it the global stabilization with guaranteed safety}.
	
	As discussed briefly in the Introduction, analyzing the robustness of systems stability in the presence of an (external) input signal can be done using the input-to-state stability (ISS) framework \cite{Sontag1989,Sontag1996}. Let us briefly recall the ISS concept from \cite{Sontag1996}. 
	
	The system \rfb{NL} is called {\em input-to-state stable} if there exist a $\beta\in\mathcal{KL}$ and $\gamma\in\mathcal{K}$ such that for any $u\in L^\infty$ and $x_0\in \mathcal{X}_0$, the following inequality holds for all $t$:
	\begin{equation}\label{iss}
		\|x(t)\|\leq\beta(\|x_0\|,t)+\gamma(\|u\|_{L^\infty([0,t))}).
	\end{equation}
	In this notion, the functions $\beta$ and $\gamma$ in \rfb{iss} describe the decaying effect from a non-zero initial condition $x_0$ and the influence of a bounded input signal $u$ to the state trajectory $x$, respectively. The Lyapunov characterization of ISS systems is provided in the following well-known theorem from \cite{Sontag1989,Sontag1996}.
	
	\begin{theorem}\label{thm1}
		The system \rfb{NL} is ISS if and only if there exists a smooth $V:\rline^n\to\rline_+$, functions $\alpha_1,\alpha_2,\alpha_3 \in \mathcal{K}_\infty$ and a function $\gamma\in\mathcal K$ such that 
		\begin{equation}\label{ISS_eq1}
			\alpha_1(\|\xi\|) \leq V(\xi) \leq \alpha_2(\|\xi\|)
		\end{equation}
		and
		\begin{equation}\label{ISS_eq2}
			\frac{\partial V(\xi)}{\partial \xi} \left(f(\xi)+g(\xi)v\right) \leq -\alpha_3(\|\xi\|) + \gamma(\|v\|)
		\end{equation}
		hold for all $\xi\in\rline^n$ and for all $v\in\rline^m$.
	\end{theorem}
	
	The notion of ISS and its Lyapunov characterization as above have been seminal in the study of nonlinear systems robustness with respect to the uncertainties in the initial conditions and to the external disturbance signals. For instance, a well-known nonlinear small-gain theorem in \cite{Jiang1994} is based on the use of $\beta$ and $\gamma$. The study of convergence input convergence state property as in \cite{Jayawardhana2010} is based on the use of ISS Lyapunov function. However, as mentioned in the Introduction, existing results on robustness have focused on the systems' stability and there is not many attention on the robustness analysis on systems' safety. 
	
	Let us recall few main results in literature on safety analysis. In order to verify the safety of system \rfb{NL} with respect to a given unsafe set $\mathcal{D}$, a Lyapunov-like function which is called barrier certificate has been introduced in \cite{Prajna2004} where the safety of the system can be verified through the satisfaction of a Lyapunov-like inequality without having to explicitly evaluate all possible systems' trajectories. \startmodif Such barrier certificate is a reminiscent of Chetaev function for analyzing instability of nonlinear systems. While the Chetaev instability theorem is used to show that the trajectory of an autonomous system always escapes any compact set, the barrier certificate is mainly applied to show that a trajectory does not enter a given compact set. \stopmodif The barrier certificate theorem is summarized in following theorem.
	
	\vspace{0.2cm}
	\begin{theorem}\label{bcthm}
		Consider the (autonomous) system \rfb{NL} with $u=0$, i.e., $\dot{x}=f(x)$ where $x(t)\in\mathcal X\subset\rline^n$, with a given unsafe set $\mathcal{D}\subset \mathcal X$ and set of initial conditions $\mathcal{X}_0\subset \mathcal X$. Assume that there exists a barrier certificate $B:\mathcal{X}\rightarrow\mathbb{R}$ satisfying
		\begin{align}
			\label{bc1}B(\xi)>0 &\quad\forall \xi\in \mathcal{D}\\
			\label{bc2}B(\xi) < 0 &\quad \forall \xi\in\mathcal{X}_0\\
			\label{nonstrictB}\frac{\partial B(\xi)}{\partial \xi} f(\xi)  \leq 0 &\quad \forall \xi\in{\mathcal{X}}\quad\text{such that}\quad B(\xi)=0.   
		\end{align}
		Then the system is safe.
	\end{theorem}
	\vspace{0.2cm}

	The proof of this theorem is based on the fact that the evolution of $B$ starting from a non-positive value (c.f. \rfb{bc2}) will never cross the zero level set due to \rfb{nonstrictB}, i.e., the state trajectory will always be safe according to \rfb{bc1}.

	Although the safety result as in Theorem \ref{bcthm} is formulated only for autonomous systems, an extension to the non-autonomous case has also been presented in \cite{Prajna2004}. For the case where an external input $u$ is considered, e.g., the complete system as in \rfb{NL}, the safety condition \rfb{nonstrictB} becomes 
	\begin{equation}\label{nonunf_barrier}
		\frac{\partial B(\xi)}{\partial \xi}\left(f(\xi) + g(\xi)v\right)\leq 0 \quad\forall (\xi,v)\in{\mathcal{X}}\times{\mathcal{U}}
	\end{equation} 
	where $\mathcal{U}\subset \rline^m$ denotes the admissible set of input. However, the condition \rfb{nonunf_barrier} is a very restrictive assumption since it must hold for all $u(t)\in \mathcal U$ including the case when the initial condition $x(0)$ is very close to $\mathcal D$. It means that when we start very close to the unsafe state, the system must always remain safe for whatever type of input signals $u$ as long as it has values in $\mathcal U$. In this case, we can say that such system is very robust with respect to bounded external input signals. In practice, we should expect a certain degree of fragility in the system, in the sense that, if we start very close to the unsafe state, a small external input signal can already jeopardize the systems' safety; a feature that is not  captured in \rfb{nonunf_barrier}.     
	
	Instead of considering the inequality \rfb{nonunf_barrier}, we will consider a more restrictive condition on $B$ for our main results later, where the non-increasing assumption of $B$ as in \rfb{nonstrictB} is replaced by a strict inequality as follows 
	\begin{equation}\label{strictB}
		\frac{\partial B(\xi)}{\partial \xi} f(\xi) \leq - \alpha(|x|_{\mathcal D})
	\end{equation}
	where $\alpha$ is a $\mathcal K$ function. 
	
	In \cite{Romdlony2016a,Wieland2007}, the use of such barrier function $B$ for control design that guarantees safety has been presented. It is shown in these works that the standard Lyapunov-based control design can directly be extended to solving the safety problem by replacing the Lyapunov function with the barrier one. Interested readers are referred to \cite{Romdlony2016a} for control design methods that solve the stabilization with guaranteed safety by merging the control Lyapunov function with the control barrier function.

	\section{Input-to-state safety}
	
	In this section, we will explore a new notion of input-to-state safety as a tool to analyze the robustness of systems' safety. In particular, we focus our study on extending existing results on barrier certificate to the input-to-state safety framework; akin to the role of Lyapunov stability theory in the input-to-state stability results.  
	
	\vspace{0.2cm}
	\begin{definition}\label{isfdef}
    The system \rfb{NL} is called {\em input-to-state safe} (ISSf) \startmodif locally in $\mathcal X\subset\rline^n$ and \stopmodif with respect to the set of unsafe state $\mathcal{D}\subset\mathcal X$ \startmodif if for all $x_0\in \rline^n\backslash \mathcal D$, there exist  $\sigma,\phi\in\mathcal{K}$, $\mu\in\mathcal{KK}$ and $\delta>0$ such that 
		\begin{equation}\label{isf}
			\sigma(|x(t)|_{\mathcal{D}}) \geq \min\{ \mu(|x_0|_{\mathcal{D}}, t),\delta\} -\phi\left(\|u(t)\|\right) 
		\end{equation}
		holds for almost all $t\in [0,\infty)$ and for all admissible\footnote{By admissible $(x_0,u)$, we mean that the tuple is such that the RHS of \rfb{isf} is strictly positive for almost all $t\geq 0$.} $(x_0,u)$, where the constant $\delta>0$ can be dependent on boundary of $\mathcal X$. \stopmodif 
	\end{definition}
	\vspace{0.2cm}
		
\startmodif If a system is ISSf, \stopmodif we can infer from \rfb{isf} that the system \rfb{NL} may be brought to the unsafe state if the $L^\infty$-norm of $u$ is sufficiently large such that the RHS of \rfb{isf} is negative. Hence one can quantify the robustness of the system's safety with respect to an external input signal using this notion. For instance, if the initial condition $x_0$ is in the neighborhood of the boundary of unsafe state $\mathcal{D}$ then \rfb{isf} shows that a small external input signal $u$ may steer the state trajectory to enter $\mathcal D$; even when the autonomous case is safe. Since the first element on the RHS of \rfb{isf} is a $\mathcal{KK}$ function, it implies that the distance between $x(t)$ and $\mathcal D$ \startmodif is lower-bounded by a strictly increasing function until $x(t)$ leaves $\mathcal X$. As this lower-bound of the distance is non-decreasing with time, \rfb{isf} means that the system can eventually withstand larger input signal.    
	
We can also take a different view to the  ISSf inequality above. 
If $u$ is considered to be a disturbance signal with known magnitude, e.g., $\|u\|_{L^\infty}\leq k$ with $k>0$, then \rfb{isf} provides us with information on the admissible $x_0$ such that the RHS of \rfb{isf} remains positive so that the system under such external disturbance will remain safe.  
	
	Let us now investigate the ISS-Lyapunov like condition for input-to-state safety of system \rfb{NL}. 
	\begin{proposition}\label{ISSf_NL_pro}
		Consider system \rfb{NL} with a given unsafe set $\mathcal{D}\subset\mathcal{X}\subset\mathbb{R}^n$. Suppose that there exists an ISSf barrier function $B\in\mathcal C^1 (\mathcal{R}^n,\mathbb{R})$ satisfying \startmodif
		\begin{align}
			\label{Bn1}-\alpha_1(|\xi|_\mathcal{D})\leq B(\xi) &\leq -\alpha_2(|\xi|_\mathcal{D}) \, \, \, \forall \xi\in\rline^n\backslash \mathcal D \\
			\nonumber \frac{\partial B(\xi)}{\partial \xi}(f(\xi)+g(\xi)v)&\leq -\alpha_3(|\xi|_\mathcal{D})+\alpha_4(\|v\|) \\
            \label{Bn2}& \qquad \qquad \forall \xi\in\mathcal{X}\backslash \mathcal D, \forall v\in \mathcal{U},
		\end{align} \stopmodif
where $\alpha_i\in\mathcal{K}_\infty$, i=1,..4.  Assume further that the system is ISS. 
    
Then the system is input-to-state safe \startmodif locally in $\mathcal X$ and w.r.t. $\mathcal{D}$. In particular,  for any $\theta,\epsilon\in (0,1)$ and for all $x_0\in \mathcal \rline^n\backslash \mathcal D$, the ISSf inequality \rfb{isf} 
 holds for all $t\geq 0$ and for all admissible $(x_0,u)$ where $\sigma(s) = s$,  $\delta = \min\{\epsilon |\xi|_{\mathcal D}\, : \, \forall\xi \in \partial \mathcal X\}$, 
\[
\mu(s,t)=\epsilon\alpha_1^{-1}(\tilde{\alpha}(\alpha_2(s),t)) \qquad \forall s,t \geq 0
\] 
and \[
\phi(s)=\alpha_2^{-1}\circ\alpha_1\circ\alpha_3^{-1}\circ\frac{\alpha_4(s)}{\theta} \qquad \forall s\geq 0
\]\stopmodif 
with $\tilde{\alpha}\in \mathcal{KK}$ be the solution of the following initial value problem
    \begin{equation*}
    \dot{y}=(1-\theta)\alpha_3\circ\alpha_1^{-1}(y), \quad y(0)=s\in \mathbb{R}_+,
    \end{equation*}
so that $\tilde{\alpha}(s,t):=y(t)$ for all $s\geq 0$. 		
	\end{proposition}
    
\startmodif 
Prior to proving this proposition, a few remarks can be made on the relation between the ISSf barrier function satisfying  \rfb{Bn1}-\rfb{Bn2} and the barrier certificate satisfying \rfb{bc1}-\rfb{nonstrictB}. First, it is easy to see that the condition \rfb{Bn1} implies \rfb{bc2} where $\mathcal X_0$ in \rfb{bc2} is $\rline^n\backslash \bar{\mathcal D}$. Second, when we consider the autonomous case (i.e., $u=0$), then \rfb{Bn2} implies the strict version of \rfb{nonstrictB} (c.f., \rfb{strictB}).  
\stopmodif
    
	\begin{proof}
    \startmodif Let us first evaluate the solution $x(t)$ of \rfb{NL} with $x_0\in\mathcal X\backslash \mathcal D$. \stopmodif From \rfb{Bn1} it follows that $|x(t)|_\mathcal{D} \geq \alpha_1^{-1}(-B(x(t)))$, thus evaluating the time derivative of $B(x(t))$ gives us  
\startmodif		\begin{align}
			\nonumber \dot{B}(x(t))&\leq -\alpha_3\circ\alpha_1^{-1}(-B(x(t)))+\alpha_4(\|u(t)\|)\\
			\nonumber &=-(1-\theta)\alpha_3\circ\alpha_1^{-1}(-B(x(t)))\\
			\label{Bdot}&\quad-\theta\alpha_3\circ\alpha_1^{-1}(-B(x(t)))+\alpha_4(\|u(t)\|),
		\end{align}    
         with $\theta\in(0,1)$ which holds whenever $x(t)\in \mathcal X\backslash \mathcal D$. \stopmodif
        
        Thus for almost all $t$ such that \startmodif $\|u(t)\|\leq\alpha_4^{-1}\circ\theta\alpha_3\circ\alpha_1^{-1}(-B(x(t)))=:\rho(x(t))$, \stopmodif inequality \rfb{Bdot} implies that 
		\begin{equation*}
			\dot{B}(x(t))\leq-(1-\theta)\alpha_3\circ\alpha_1^{-1}(-B(x(t)))
		\end{equation*}
\startmodif holds whenever  $x(t)\in \mathcal X\backslash \mathcal D$. \stopmodif By letting $\tilde{B}(x(t))=-B(x(t))$, the last inequality becomes 
		\begin{equation}\label{initproblem1}
			\dot{\tilde{B}}(x(t))\geq(1-\theta)\alpha_3\circ\alpha_1^{-1}(\tilde{B}(x(t))).
		\end{equation}
		Note that function $(1-\theta)\alpha_3\circ\alpha_1^{-1}(r)$ belongs to $\mathcal{K}$ function and the function $\tilde{B}$ is positive definite. Hence, the RHS of \rfb{initproblem1} is always positive. Now by the comparison lemma,
		\begin{equation}\label{initproblem2}
			\tilde{B}(x(t))\geq\tilde{\alpha}(\tilde{B}(x_{0}),t)
		\end{equation} 
		where $\tilde{\alpha}\in\mathcal{KK}$ is the solution $y(t)$ of  
		\begin{equation*}
			\dot{y}=(1-\theta)\alpha_3\circ\alpha_1^{-1}(y), \quad y(0)=s\in \mathbb{R}_+,  
		\end{equation*}
        i.e., $\tilde{\alpha}(s,t):=y(t)$ for any positive initial condition $s$.
        
		By subtituting \rfb{initproblem2} into the lower bound and upper bound of $B(x)$ in \rfb{Bn1} it follows that 
		\begin{align}
			\nonumber &\alpha_1(|x(t)|_\mathcal{D})\geq\tilde{\alpha}(\tilde{B}(x_0),t) \geq \tilde{\alpha}(\alpha_2(|x_0|_\mathcal{D}),t)\\
			&\implies |x(t)|_\mathcal{D}\geq \label{KK_init}\alpha_1^{-1}\tilde{\alpha}(\alpha_2(|x_0|_\mathcal{D}),t) =: \widetilde\mu(|x_0|_\mathcal{D},t)
		\end{align}  
\startmodif	which holds for almost all	$t$ s.t. $\|u(t)\|\leq \rho(x(t))$ and whenever $x(t)\in \mathcal X\backslash \mathcal D$. \stopmodif
    
		Now, let us consider the other case where $\|u(t)\| > \rho(x(t))$. In this case, it follows immediately that
		\begin{align}
			\nonumber&-B(x(t))\leq \alpha_1\circ\alpha_3^{-1}\circ\frac{\alpha_4(\|u(t)\|)}{\theta}\\
			\nonumber&\implies\alpha_2(|x(t)|_\mathcal{D})\leq \alpha_1\circ\alpha_3^{-1}\circ\frac{\alpha_4(\|u(t)\|)}{\theta}\\
			\label{K_u}&\implies|x(t)|_\mathcal{D}\leq \alpha_2^{-1}\circ\alpha_1\circ\alpha_3^{-1}\circ\frac{\alpha_4(\|u(t)\|)}{\theta} =: \widetilde\phi(\|u(t)\|)
		\end{align}
        
\startmodif 
We will now combine these two cases as follows. Firstly, from \rfb{KK_init}, it follows that 
\begin{multline}\label{KK_init_1}
-\epsilon \widetilde\mu(|x_0|_\mathcal{D},t) + |x(t)|_\mathcal{D} \\ \geq (1-\epsilon) \widetilde\mu(|x_0|_\mathcal{D},t) - \eta \widetilde\phi(\|u(t)\|),
\end{multline}
where $\epsilon, \eta \in (0,1)$.  
This inequality is obtained by adding both sides of \rfb{KK_init} by $-\epsilon \widetilde\mu(|x_0|_\mathcal{D},t)$ and substracting the right-hand side of \rfb{KK_init} by $- \eta \widetilde\phi(\|u(t)\|)$ which is non-positive for all $u(t)$. On the other hand, by multiplying both sides of \rfb{K_u} by $-\eta$ and then by adding both sides by $(1-\epsilon) \widetilde\mu(|x_0|_\mathcal{D},t)$, we get
\begin{multline}\label{K_u_1}
(1-\epsilon) \widetilde\mu(|x_0|_\mathcal{D},t) -\eta |x(t)|_\mathcal{D} \\ \geq (1-\epsilon) \widetilde\mu(|x_0|_\mathcal{D},t) - \eta \widetilde\phi(\|u(t)\|).
\end{multline}
Thus, \rfb{KK_init_1} (which holds for $\|u(t)\| \leq \rho(x(t))$) and \rfb{K_u_1} (which is true for $\|u(t)\| > \rho(x(t))$) imply that
\begin{multline}\label{K_combined}
\max\left\{\bluff-\epsilon \widetilde\mu(|x_0|_\mathcal{D},t) + |x(t)|_\mathcal{D}, \right. \\ \left. \bluff (1-\epsilon) \widetilde\mu(|x_0|_\mathcal{D},t) -\eta |x(t)|_\mathcal{D} \right\} \\ \geq (1-\epsilon) \widetilde\mu(|x_0|_\mathcal{D},t) - \eta \widetilde\phi(\|u(t)\|)
\end{multline}
holds for all $t\geq 0$ s.t. $x(t)\in\mathcal X\backslash\mathcal D$. 

Since the state trajectory starts from the safe region, then for a given initial condition $x_0$ and bounded input $u$, there exists sufficiently small $\eta,$ $\epsilon$ and $T_1>0$ such that the right hand side of \rfb{K_combined} and each term on the left-hand side are  positive for all $t\in [0,T_1)$. Thus, since $\max\{a,b\}\leq a+b$ for $a,b \geq 0$, \rfb{K_combined} implies that
\begin{align}
\nonumber & (1-2\epsilon) \widetilde\mu(|x_0|_\mathcal{D},t) + (1-\eta) |x(t)|_\mathcal{D} \\
\nonumber & \qquad \geq (1-\epsilon) \widetilde\mu(|x_0|_\mathcal{D},t) - \eta \widetilde\phi(\|u(t)\|) \\
\nonumber \Leftrightarrow & (1-\eta) |x(t)|_\mathcal{D} \geq \epsilon \widetilde\mu(|x_0|_\mathcal{D},t) - \eta \widetilde\phi(\|u(t)\|) \\
\label{ISSf_eq1}\Leftrightarrow & |x(t)|_\mathcal{D} \geq \frac{\epsilon}{1-\eta} \widetilde\mu(|x_0|_\mathcal{D},t) - \frac{\eta}{1-\eta} \widetilde\phi(\|u(t)\|)
\end{align}
holds for almost all $t\in [0,T_1)$. 

We will prove now that we can extend the time interval, where \rfb{ISSf_eq1} is valid, to $[0,T_{1,\max})$ with finite $T_{1,\max}<\infty$ if $x$ leaves the set $\mathcal X$ at time $T_{1,\max}$, or $T_{1,\max}=\infty$ when $x$ stays in $\mathcal X\backslash \mathcal D$ at all time. In particular, we show that we can choose $\eta$ and $\epsilon$ such that both terms on the LHS of \rfb{K_combined} are positive for almost all  $t\in [0,T_{1,\max})$, so that \rfb{ISSf_eq1} holds accordingly. 

Firstly, let us show that for any $\epsilon \in (0,1)$, there exists $\eta \in (0,1)$ such that 
\begin{equation}\label{condition_1}
|x(t)|_{\mathcal D} \leq \frac{1-\epsilon}{\eta}\widetilde\mu(|x_0|_{\mathcal D},t) \qquad \forall t\in[0,\infty).
\end{equation}
Since the system is ISS, there exists $\beta\in\mathcal{KL}$ and $\gamma\in\mathcal{K}_\infty$ such that 
\begin{align*}
|x(t)| & \leq \beta(|x_0|,t) + \gamma(\|u\|_{L^\infty}) \\
& \leq \beta(|x_0|,0) + \gamma(\|u\|_{L^\infty}) =: D_1.
\end{align*}
By triangular inequality and by denoting $D_2=\max\{|\xi | \ : \ \forall\xi\in\mathcal D\}$, it follows that
\begin{align}
\nonumber|x(t)|_{\mathcal D} & \leq D_2 + |x(t)| \leq D_1 + D_2 \\
\label{condition1_ineq_a}& \leq \frac{D_1+D_2}{\widetilde\mu (|x_0|_{\mathcal D},0)} \widetilde\mu (|x_0|_{\mathcal D},t),
\end{align}
where the last inequality is due to the fact that $\widetilde\mu (|x_0|_{\mathcal D},t)\geq \widetilde\mu (|x_0|_{\mathcal D},0)$ for all $t\geq 0$. Thus, by taking 
\begin{equation}\label{eta_eq}
\eta = \min \left\{0.5, \frac{(1-\epsilon)\widetilde\mu (|x_0|_{\mathcal D},0)}{D_1+D_2}\right\} \in (0,0.5],
\end{equation}
the inequality \rfb{condition1_ineq_a} implies that \rfb{condition_1} holds for all $t\geq 0$. Hence, the second term on the LHS of \rfb{K_combined} is always positive for all $t$. 

It remains now to check whether 
\[
|x(t)|_{\mathcal D} > \epsilon \widetilde\mu (|x_0|_{\mathcal D},t)
\]
for all $t\in [0,T_{1,\max})$. We will show this by contradiction. Suppose that there is a finite $\tau<T_{1,\max}$ that defines the time when $|x(\tau)|_\mathcal{D} = \epsilon \widetilde\mu (|x_0|_{\mathcal D},\tau)$. In this case, \rfb{ISSf_eq1} still holds and we have that
\begin{align*}
|x(\tau)|_\mathcal{D} & \geq \frac{\epsilon}{1-\eta} \widetilde\mu(|x_0|_\mathcal{D},\tau) - \frac{\eta}{1-\eta} \widetilde\phi(\|u(\tau)\|) \\ 
& = \epsilon \widetilde\mu(|x_0|_\mathcal{D},\tau) \\& \qquad + \frac{\eta}{1-\eta}\left( \epsilon\widetilde\mu(|x_0|_\mathcal{D},\tau) -\widetilde\phi(\|u(\tau)\|)\right).
\end{align*}
Since $\widetilde\phi(\|u(t)\|) < \epsilon\widetilde\mu(|x_0|_\mathcal{D},t)$ for all $t\geq 0$ (by hypothesis of the proposition on the admissibility of $(x_0,u)$ with $\mu = \epsilon\widetilde\mu$ and $\widetilde\phi = \phi$), it follows from the above inequality that
\[
|x(\tau)|_{\mathcal D} > \epsilon \widetilde\mu(|x_0|_\mathcal{D},\tau) 
\]
which is a contradiction. Thus, we have that \rfb{ISSf_eq1} holds for almost all $t\in [0,T_{1,\max})$. 

Finally, we will derive the conservative lower bound of \rfb{ISSf_eq1} such that it will no longer depend on $\eta$ (which is currently dependent on $x_0$ and $u$ as in \rfb{eta_eq}). By the definition of $\eta$ in \rfb{eta_eq}, it is trivial to check that $0< \eta < 0.5$,  
\[
1 < \frac{1}{1-\eta} < 2 \ \ \text{and} \ \ 0 > \frac{-\eta}{1-\eta} > -1.
\]
Thus, \rfb{ISSf_eq1} implies that 
\begin{align}
\label{withinX_eq}|x(t)|_\mathcal{D} & \geq \epsilon \widetilde\mu(|x_0|_\mathcal{D},t) -  \widetilde\phi(\|u(t)\|) 
\end{align}
for almost all $t\in [0,T_{1,\max})$.

On the other hand, by defining $\kappa:=\min\{|\xi|_{\mathcal D}\,:\,\forall\xi\in\partial \mathcal X\} > 0$, we have that when $x(t)\notin \mathcal X$ (including for the second case when $x_0\notin\mathcal X$),
\begin{equation}\label{outsideX_eq}
|x(t)|_{\mathcal D} \geq \kappa \geq \kappa - \widetilde\phi(\|u(t)\|).
\end{equation}
Once $x$ leaves $\mathcal X$ and enters again $\mathcal X$ at a later time interval, then we can use again the argument as before where the initial condition is taken in the neighborhood of the boundary of $\mathcal X$. Indeed, suppose that $x$ enters again $\mathcal X$ at time $T_2>T_{1,\max}$. Then by following the same argument as before, we get
\begin{align}
\nonumber |x(t)|_\mathcal{D} & \geq \epsilon \widetilde\mu(|x(T_2)|_\mathcal{D},t-T_2) -  \widetilde\phi(\|u(t)\|) \\
\label{enterX_eq} & \geq \epsilon \widetilde\mu(\kappa,0) -  \widetilde\phi(\|u(t)\|),
\end{align}
for almost all $t\in [T_2,T_{2,\max})$ where $T_{2,\max}$ is the maximum time where $x$ remains in $\mathcal X$.

Since in all of these cases, $|x(t)|_{\mathcal D}$ satisfies either \rfb{withinX_eq}, \rfb{outsideX_eq} or \rfb{enterX_eq} in different time intervals, we can combine them by taking the minimum of their lower bounds. Thus by defining $\delta:=\epsilon\widetilde\mu(\kappa,0)$ with $\kappa$ as defined before \rfb{outsideX_eq},  
\begin{align*}
|x(t)|_{\mathcal D} & \geq \min\{\epsilon \widetilde\mu(|x_0|_\mathcal{D},t)\, , \, \kappa \, , \, \epsilon\widetilde\mu(\kappa,0)\} -  \widetilde\phi(\|u(t)\|) \\
& = \min\{\epsilon \widetilde\mu(|x_0|_\mathcal{D},t)\, , \delta\} -  \widetilde\phi(\|u(t)\|)
\end{align*}
holds for almost all $t\in [0,\infty)$. 

Hence, we have ISSf with $\mu = \epsilon \widetilde \mu$ and $\phi = \widetilde\phi$ where $\widetilde \mu$ and $\widetilde \phi$ are as in \rfb{KK_init} and \rfb{K_u}, respectively, and $\delta$ as defined above. Note that the choice of $\epsilon\in (0,1)$ is, in this case, independent of admissible tuple $(x_0,u)$.   
\stopmodif

	\end{proof}
    
\startmodif The ISS assumption in this proposition can be relaxed by weaker conditions that can guarantee the boundedness of $|x(t)|_{\mathcal D}$ so that the inequality \rfb{condition_1} in the proof of Proposition \ref{ISSf_NL_pro} holds. For instance, we can assume that the system is integral input-to-state stable or it is practically input-to-state stable. \stopmodif
    
	   One can see from Proposition \ref{ISSf_NL_pro} that the inequalities in \rfb{Bn1} and \rfb{Bn2} are reminiscent to those used in the study of ISS Lyapunov function. In this context, the inequality \rfb{Bn2} resembles the dissipation inequality in the ISS Lyapunov function and the growth of $B$ as in \rfb{Bn1} can be likened to the growth of $V$ as in \rfb{ISS_eq1}, albeit they grow with different sign as well as with different metric norm.

    We can now combine the notion of input-to-state stability and that of input-to-state safety which allows us to study the robustness of a stable and safe system with respect to an external input signal $u$.
	
	\vspace{0.2cm}
	\begin{definition}
		System \rfb{NL} is called ISS with guaranteed safety (ISS-GS) with respect to $\mathcal{D}$ \startmodif if there exists $\mathcal X\subset \rline^n$ such that the system \rfb{NL} is both input-to-state stable 
		and input-to-state safe locally in $\mathcal X$ and w.r.t. $\mathcal{D}\subset \mathcal X$.  \stopmodif
	\end{definition}
	\vspace{0.2cm}

	It is trivial to show that if there exist both an  ISS Lyapunov function $V$ satisfying \rfb{ISS_eq1}--\rfb{ISS_eq2} and an ISSf barrier function $B$ satisfying \rfb{Bn1}--\rfb{Bn2} locally on $\mathcal{X} \subset \rline^n$ with $\mathcal D\subset \mathcal{X}$ then the system is input-to-state stable with guaranteed safety. Instead of considering two separate functions $V$ and $B$ as suggested before, we can also consider combining the ISS Lyapunov inequality \rfb{ISS_eq2} and ISSf barrier inequality \rfb{Bn2} as shown in the following proposition.

	\begin{corollary}\label{Wprop}
		Suppose that there exists $W:\rline^n\to\rline$ and $\mathcal D\subset \mathcal{X} \subset \rline^n$ such that \startmodif
		\begin{align}
		\label{Wn1} \alpha_1(\|\xi\|)\leq W(\xi) &\leq \alpha_2(\|\xi\|) \quad \forall \xi\in\mathbb{R}^n \\
		\label{Wn2}-\alpha_3(|\xi|_\mathcal{D})\leq W(\xi)-c&\leq-\alpha_4(|\xi|_\mathcal{D}) \quad \forall \xi\in\mathcal{X}\backslash \mathcal D\\
		\nonumber\frac{\partial W(\xi)}{\partial \xi}(f(\xi)+g(\xi)v)&\leq-\alpha_5(\|\xi\|)-\Xi_{\mathcal{X}}(\xi)\alpha_6(|\xi|_\mathcal{D}) \\
      \label{Wn3}  & \ \ \ \  +\alpha_7(\|v\|)
		\end{align} \stopmodif
		where $\Xi_{\mathcal{X}}$ is an indicator function for $\mathcal{X}$, $c>0$, the functions $\alpha_i \in \mathcal{K}_\infty $ for $i=1,..7$. Then it is ISS with guaranteed safety with respect to $\mathcal{D}$. 
	\end{corollary}
	\vspace{0.2cm}
	
	\begin{proof}
		It is trivial to check that $W(x)$ qualifies as an ISS Lyapunov function satisfying \rfb{ISS_eq1}--\rfb{ISS_eq2} and as an ISSf barrier function satisfying \rfb{Bn1}--\rfb{Bn2} locally in $\mathcal X$. \startmodif The ISS property follows trivially from \rfb{Wn1} and \rfb{Wn3} and Theorem \ref{thm1}. 
        
        Let $B(\xi) = W(\xi) -c$ for all $\xi \in \mathcal X\backslash \mathcal D$. Subsequently, let the function $B$ be extended smoothly to $\xi\in\rline^n\backslash \mathcal X$ so that \rfb{Bn1} holds for all $\rline^n\backslash\mathcal D$. It follows from \rfb{Wn3} that  
		\[
		\frac{\partial B(\xi)}{\partial \xi}(f(\xi)+g(\xi)v) \leq -\alpha_6(|\xi|_{\mathcal D})+\alpha_7(\|v\|)
		\]
        holds for all $\xi\in\mathcal{X}\backslash \mathcal D$ and for all $v\in \mathcal{U}$. By Proposition \ref{ISSf_NL_pro}, it implies that 
it is ISSf. \stopmodif
	\end{proof}

	\section{Simulation result}
	In this section, we consider an example of a simple mobile robot navigation described by the following equations
	\begin{align}\label{robot}
		\nonumber\dot{x}_1 &=v_1+u_1\\
		\dot{x}_2 &=v_2+u_2
	\end{align}
	where $x=[x_1,x_2]^T$ is the position in a 2D plane, $v=[v_1,v_2]^T$ is its velocity which is used as a feedback control input, and $u=[u_1,u_2]^T\in L^\infty$ is a bounded disturbance signal.

\startmodif
We assume that the unsafe state domain is given by $\mathcal{D}:=\{x\in\mathbb{R}^2|(x_1-4)^2+(x_2-6)^2<4\}$ 
and consider a bounded disturbance signal $u$ where for numerical purposes its $L_\infty$-norm is bounded by $3$.  
 We are now interested in designing a control law such the closed-loop system is ISS with guaranteed safety. 
 
It is straightforward to check that the system \rfb{robot} can be made ISS by applying the control law $\sbm{v_1\\v_2}=-\nabla_x V(x)$ with $V(x)=x_1^2+x_1x_2+x_2^2$.  
On the other hand, one can evaluate that the control law $\sbm{v_1\\v_2}=-\nabla_x B(x)$ with $B(x)=-(x_1-4)^2-(x_2-6)^2+4$ ensures that the closed-loop system is ISSf. 

We will now try to combine both control laws following the same construction as in \cite{Romdlony2016a} where a control Lyapunov function and a control barrier function can be combined. 

Firstly, we will modify $B$ such that it will be a compactly-supported function in the neighborhood of $\mathcal D$, which will later be related to the set $\mathcal X$ in \rfb{Wn2} and \rfb{Wn3} in Corollary \ref{Wprop}. For numerical simulation, we will consider $\mathcal{X}:=\mathcal{D}+\mathbb{B}_{0.5}(0)=\{x\in\mathbb{R}^2|(x_1-4)^2+(x_2-6)^2<9\}$. Using the above $B$, we can define a compactly-support function $\widetilde B$ as follows \cite{Romdlony2016a}.
		\begin{multline*}
			\widetilde{B}(x)=B(\omega)+\oint\limits_{\Gamma} 0.5\left(\text{cos}\left(\frac{\pi}{\delta}B(\sigma)\right)+1\right)\frac{\partial B(\sigma)}{\partial x} d\sigma \\ \forall x \in\mathcal{X}
		\end{multline*} 
		where $\omega\in\partial\mathcal{D}$ is any point in the boundary of $\mathcal D$, $\Gamma$ is any path from point $\omega$ to any point $\phi \in\mathcal{X}$, and $\delta=-B(\partial\mathcal{X})=5$. For $x\in\mathbb{R}^2\setminus\mathcal{X}$, $\widetilde{B}(x)$ is defined as $-\delta$. 
		
\begin{figure}
		\centering
		\includegraphics[height=2.5in, width=3.5in]{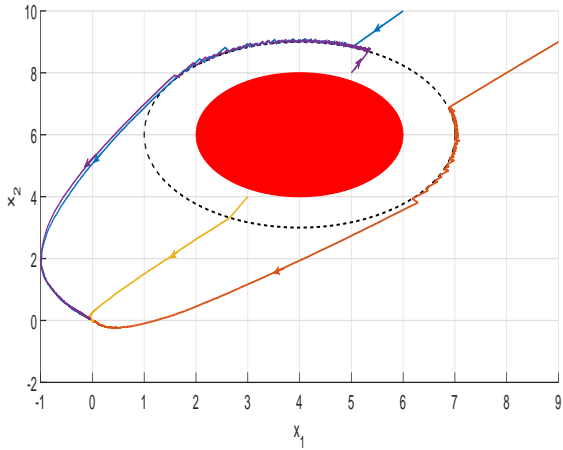}
		\caption{State trajectories $x(t)$ starting from four different initial conditions. The set of unsafe state $\mathcal{D}$ is shown in red area, and the boundary of $\mathcal{X}$ is shown by dashed line.}
		\label{x1-x2_ISS-GS}
	\end{figure}

\begin{figure}
		\centering
		\includegraphics[height=3in, width=3.5in]{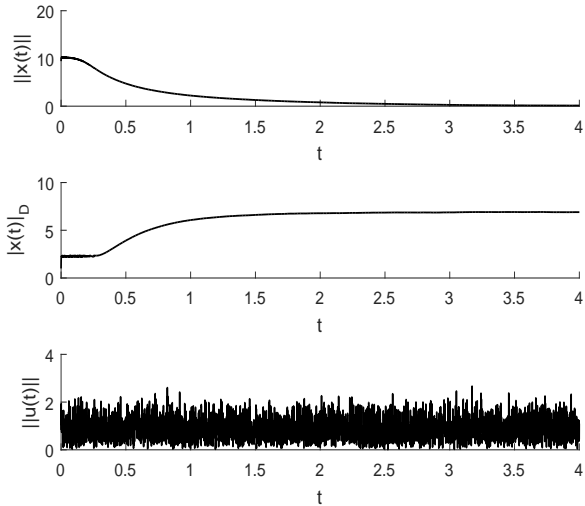}
		\caption{The time plots of $\|x(t)\|$ and $|x(t)|_\mathcal{D}$ started from $x_0=(5,8)$, and disturbance signal $u(t)$.} 
		\label{time_ISS-GS}
	\end{figure}

		Following the same procedure discussed in \cite{Romdlony2016a}, 
we can merge both $V$ and $\tilde B$ into $W(x) = V(x)+k_1 \widetilde{B}(x)+k_2$ where we can choose $k_1=100$ and $k_2=-10$ such that \rfb{Wn1}-\rfb{Wn3} hold. Using $W$, the control law which achieves ISS with guaranteed safety of the closed-loop system is given by the gradient control law $v=-\nabla_x W(x)=-\frac{\partial^T W}{\partial x}$. Its explicit formula is given by \stopmodif
		\begin{equation}
			v=\left\{\begin{array}{ll}
				-\nabla_x V(x)-k_1\nabla_x \widetilde{B}(x) & \forall x \in \mathcal{X}\\
				-\nabla_x V(x) & \forall x \in \mathbb{R}^2\setminus\mathcal{X}.
			\end{array}\right.
		\end{equation}  		
		Figure \ref{x1-x2_ISS-GS} shows the evolution of state $x_1$ and $x_2$ starting from four different initial conditions. \startmodif It can be seen from the figure that all state trajectories of the closed-loop system converge to origin while avoiding the unsafe state despite being perturbed by an external disturbance signal $u$. When we evaluate the evolution of $\|x(t)\|$ and $|x(t)|_\mathcal{D}$ that is started from $x_0=(5,8)$, Figure \ref{time_ISS-GS} shows clearly the ISS with guaranteed safety property of the closed-loop system. \stopmodif 
        
	\section{Conclusion}
	In this paper, we have presented a novel notion of input-to-state safety which is complementary to the well-known input-to-state stability notion. The new notion has allowed us to characterize the evolution of the state distance to the set of unsafe state whose lower bound depends on the initial condition and the external input signal. It can be used for the robustness analysis of systems' safety against external disturbances. 
	
		\section{Acknowledgment}
\startmodif We would like to thank anonymous reviewers for their suggestions and technical comments which have improved the paper. \stopmodif

\end{document}